\documentclass{emulateapj}
\newcommand{\abs}[1]{\left\vert#1\right\vert}
\newcommand{\be}{  \begin{eqnarray} }
\newcommand{\ee}{  \end{eqnarray} }

\newcommand{\icarus}{Icar}

\newcommand{\beq}{\begin{equation}}
\newcommand{\eeq}{\end{equation}}

\renewcommand{\vec}[1]{\mbox{\boldmath $\displaystyle #1$}}
\newcommand{\bxi}{{\mbox{\boldmath $\xi$}}}

\newcommand{\grad}{{\mbox{\boldmath $\nabla$}}}

\def\simless{\mathbin{\lower 3pt\hbox
      {$\rlap{\raise 5pt\hbox{$\char'074$}}\mathchar"7218$}}}
\def\simgreat{\mathbin{\lower 3pt\hbox
      {$\rlap{\raise 5pt\hbox{$\char'076$}}\mathchar"7218$}}} 

\def\spose#1{\hbox to 0pt{#1\hss}}
\def\lta{\mathrel{\spose{\lower 3pt\hbox{$\mathchar"218$}}
     \raise 2.0pt\hbox{$\mathchar"13C$}}}
\def\gta{\mathrel{\spose{\lower 3pt\hbox{$\mathchar"218$}}
     \raise 2.0pt\hbox{$\mathchar"13E$}}}
\font\gkvec=cmmib10                         
\def\bxi{\hbox{{\gkvec\char24}}}           

\begin{document}
\title{Q in other Solar Systems}
\author{Aristotle Socrates$^{*}$, Boaz Katz$^{ *\dag}$  \& Subo Dong$^{\dag\dag}$}
\affiliation{Institute for Advanced Study, Princeton, NJ 08540, USA}
\begin{abstract}
A significant fraction of the hot Jupiters  with final circularized orbital periods 
of $\lesssim 5$ days are thought to form through the channel of high-eccentricity migration. 
Tidal dissipation at successive periastron passages removes orbital 
energy of the planet, which has the potential for changes in semi-major axis
of a factor of ten to a thousand.  In the equilibrium tide approximation
we show that, in order for high-eccentricity migration to 
take place, the relative level of tidal dissipation 
in Jupiter analogues must be at least 10 times higher than the upper-limit 
attributed to the Jupiter-Io interaction.  While this is not a severe problem 
for high$-e$ migration, it contradicts the results of several previous 
calculations. We show that these calculations of high$-e$ migration inadvertently over-estimated the
strength of tidal dissipation by three to four orders of magnitude.  These discrepancies 
were obscured by the use of various parameters, 
such as lag time $\tau$, tidal quality factor $Q$ and viscous time $t_{V}$.  We provide the values
of these parameters required for the Jupiter-Io interaction, tidal circularization and 
high-$e$ migration.   Implications 
for tidal theory as well as models of the inflated radii of hot Jupiters are discussed.
Though the tidal $Q$ is not, in general,
well-defined, we derive a formula for it during high-eccentricity migration
where $Q$ is approximately constant throughout evolution.

\end{abstract}

\maketitle

\section{Introduction}

The recent explosion in the discovery of extrasolar planets has re-invigorated interest
in theory of tidal excitation and dissipation.  The proximity of the hot Jupiters
to the stellar host suggests that tidal forces shaped or are currently shaping the orbital 
and structural properties of these close-in planets. 

High-$e$ migration (e.g., Rasio \& Ford 1996; Wu \& Murray 2003; Fabrycky 
\& Tremaine; see Socrates et al. (2011) for a summary) is a class of 
migration scenarios that
depend upon the excitation of eccentricity, $e$, to high values where tidal dissipation is activated
as small periastron distances.
Successive periastron passages convert orbital energy into heat that is 
ultimately radiated away.  The discovery that the orbits of a significant fraction of
the hot Jupiters are mis-aligned with the spin of their stellar hosts lends 
strength to the claim that some high$-e$ migration is at work (Winn et al. 2010).  

For such mechanisms to work, tidal dissipation during periastron passage must be strong
enough to dissipate the required amount of orbital 
energy within the age of the system, which for FGK stars on the main 
sequence is a significant fraction of the Hubble time.  This requirement
places a lower bound on the strength of dissipation, which can then 
be tested against theories of the tidal excitation and dissipation.
By utilizing equilibrium tidal theory in the weak friction approximation,
Goldreich \& Soter (1966; hereafter GS) produced the first estimate of the level 
of dissipation in gas giants (Jupiter and Saturn), parameterized by a tidal 
quality factor $Q$.   

There is a general understanding in the literature
that high-$e$ migration is a viable mechanism for forming hot Jupiters
in the event that the relative level of tidal dissipation, within the context of 
equilibrium theory, is consistent with that inferred
by GS's study of the Jupiter-Io interaction.  Here we critically re-examine
the viability of high-$e$ migration and tidal circularization for 
hot extrasolar gas giant planets.  We do so within the context of the equilibrium
tidal theory of Hut (1981), which is widely employed in the study of high-$e$
migration (e.g., Eggleton et al. 1998; Wu\& Murray 2003; Fabrycky \& Tremaine 2007;
Naoz et al. 2011).

In \S\ref{s: theory} we summarize results from equilibrium tide theory, 
in the limit of weak dissipation, that are relevant for this work.  
We revisit the concept of the tidal quality factor
$Q$ and examine its relation to the time lag $\tau$ of the equilibrium 
tidal response.  The required $\tau$
for high$-e$ migration is estimated in \S\ref{s: calibration}.
There we show that the required strength of tidal dissipation for
high-$e$ migration is at least 10 times stronger than the inferred
upper-limit for dissipation arising from the Jupiter-Io interaction.
In \S\ref{s: previous} we briefly review some previous calculations of high$-e$ extrasolar
planet migration and comment on their prescriptions for tidal dissipation.
We also discuss  \S\ref{s: implications} 
some specific dynamical tide models in the literature and show that
they cannot account for circularization of the hot Jupiters and thus are unlikely 
account for their presumed high-$e$ migration.
We summarize in \S\ref{s: summary}.

\section{Theory of Equilibrium Tides}
\label{s: theory}

\subsection{Physical Basis for Equilibrium Tides}
\label{ss: basis}

Hut (1981) asserts that the most attractive feature of the constant
lag time model for tidal dissipation is its simplicity.  That is, it is straightforward to implement 
in analyses of secular orbital dynamics.  However, Socrates \& Katz (2012) show that 
a constant single time lag $\tau$ follows directly from the equations of motion, given 
the following basic assumptions
\begin{itemize}
\item{the tidal forcing and dissipation mechanism are slow and non-resonant\footnote{Here, `non-resonant' refers to the fluid response of the forced body.}}
\item{the equilibrium structure of the forced body is spherically 
symmetric}
\item{the dissipation is weak such that the decay time of the forced 
response is long in comparison to the characteristic forcing periods}
\item{only the quadrapolar component of the tidal force is important }
\item{non-linear effects are unimportant }
\end{itemize} 

Altogether, these conditions lead to a constant lag time $\tau$ of the forced
response, which is an intrinsic property of the tidally forced object in question.  
Therefore, if Jupiter were to be placed in another system where the 
characteristic forcing frequency and amplitude are different than that felt 
by Jupiter from Io, the lag time $\tau$ nevertheless remains the same as
long as the assumptions above are correct.

Socrates \& Katz (2012) show that, given the assumptions listed above, 
the constant lag time prescription follows directly from the equations of
motion for a spherically symmetric forced fluid body.  In the Appendix we show that, within the equilibrium tide approximation, quantification 
of the level of tidal dissipation with a tidal quality factor $Q$ is only possible
under the assumption of a single forcing frequency and a single constant lag time.  

\subsection{relevant results of equilibrium tides}

Under the simplifying assumption of spin-orbit alignment, the amount of energy dissipated per orbit
is given by
\begin{eqnarray}
\label{e: Delta_E}
\Delta  E & = - & \oint_{\rm orb}\,dt\,dE/dt\\
&= & E_{\rm F}\left(\frac{P_{\rm F}}{t_{_D}}\right)\times\left[A-2\,B\,x+C\,x^2 \right]\nonumber
\end{eqnarray}
and the change in specific orbital angular momentum per orbit is
\begin{eqnarray}
\label{e: Delta_h}
\Delta J & = & \oint_{\rm orb}\,dt\,dJ/dt \\
& =& -J\left(\frac{P_{\rm F}}{t_{_D}}\right)\times\left[B -x\,C \right]\nonumber
\end{eqnarray}
where $E_{\rm F}=GM\left(M+m_{\rm per}\right)/a_{\rm F}$, $J=\sqrt{G(M+m_{\rm per}) 
a_{\rm F}}$, $P_{\rm F}=2\pi/\sqrt{G\left(M+m_{\rm per}\right)/a^3_{\rm F}}
=2\pi/n_{\rm F}$ 
and $a_{\rm F}\equiv a\left(1-e^2\right)$. Here $x=\Omega/n_{\rm F}$  where
$\Omega$ is the rotation rate and 
$A-C$ are functions of $e$ and given by eqs. A33 - A35
of 
Hut (1981).  In the event of pseudo synchronous rotation where $\Omega=\Omega_{\rm ps}=n_{\rm F}B/C$, orbital angular momentum $J$ is conserved and
 $a_{\rm F}$ is a constant and equal to the final circularization radius. 
Here $M$ is the mass of the forced body, $m_{\rm per}$ is the mass of the perturber
and 
\be
t_D\equiv\frac{M\,a^8_{\rm F}}{3\,k_L\,\tau\,G\,m_{\rm per}^2\,R^5}
\label{e: t_D}
\ee
is a characteristic time scale for tidal dissipation, where $\tau$ is the lag time,
$k_L$ is the Love number and $R$ is the radius of the forced body.

The evolutionary equations for the semi-major axis $a$ and
eccentricity $e$ are obtained by dividing eqs. (\ref{e: Delta_E}) and (\ref{e: Delta_h})
by the orbital periods and then using Kepler's laws to relate the orbital energy $E$
and $J$ to $e$ and $a$.  The evolutionary equations
in three limiting cases of interest are provided below.

\subsubsection{synchronization:  $e=0$, $\Omega\neq\Omega_{\rm ps}$, $M\gg m_{\rm per}$}
Spin down of a major body by a minor body leads to expansion of the orbital 
separation between the two as a result of angular momentum conservation.  The rate of
this expansion is given by
\be
\frac{\dot a}{a}=-\frac{2\left(n-\Omega\right)}{n\,t_{_D}}\frac{M}{m_{\rm per}}
\label{e: adot_synch}
\ee
where $n$ is the mean motion.
The expression above determines the rate of orbital expansion e.g., of
the Moon from the Earth.  The ultimate source of energy and angular momentum
originate from the spin of the major body.

\subsubsection{circularization:  $e\ll 1$, $\Omega=\Omega_{\rm ps}$, $M\ll m_{\rm per}$}

When $e\ll1$, the pseudosynchonrous spin rate $\Omega_{\rm ps}\approx n$
and the rate of decay of orbital eccentricity follows
\be
\frac{\dot{e}}{e}=-\frac{2}{t_{_D}}.
\label{e: edot_circ}
\ee
When the orbital angular momentum is large compared to the spin angular
momentum, the forced body attains a pseudo synchronous state before 
circularization can proceed.  The source of energy for dissipation during migration
is primarily orbital.

\subsubsection{high-$e$ migration: $e\rightarrow 1$, $\Omega=\Omega_{\rm ps}$, $M\ll m_{\rm per}$}

As in the case of circularization the spin is pseudo synchronous during high-$e$
migration and therefore, the orbital angular momentum $J$ remains a constant during 
inward migration.  Consequently, the shape of the orbit near periastron is approximately
constant as long as $e\approx 1$ and therefore, the energy 
dissipated per orbit $\Delta E$ (e.g., Socrates et al. 2012) is approximately a constant
as well.  

In the limit that $e\rightarrow 1$ the 
migration rate is given by 
\be
\frac{\dot a}{a}=-\frac{2007}{64\,t_{_D}}\sqrt{\frac{a_{\rm F}}{a}}
\label{e: adot_high_e}
\ee
which is consistent with the expectation that $\Delta E\approx $ a constant
per orbit.

\subsection{Q \& $\tau$}
\label{ss: Q}

Since Goldreich \& Soter's (1966) measurement of the tidal quality 
factor $Q$ of Jupiter, $Q$ has become a common parameterization for the level
of tidal dissipation in gas giant planets and stars.  In what follows, 
we discuss the relationship and applicability of $Q$ and
and the lag time $\tau$.

In physics, the quality factor $Q$ is often utilized to parameterize
the level of dissipation in a forced damped harmonic oscillator
that obeys
\be
\ddot{X}+\omega^2\,X+\gamma\,\dot{X}=F(t)
\label{e: harm_osc}
\ee
with $F=\sqrt{\frac{1}{2}}F_0\, e^{i\sigma t}$ + c.c.  The fundamental 
frequency $\omega$ and damping rate $\gamma$ are intrinsic properties
of the oscillator. When the 
forcing is slow such that $\sigma\ll\omega$ and 
if the dissipation is weak, 
then the leading order solution $X^{(0)}$ is given by
\be
X^{(0)}=\frac{F(t)}{\omega^2}
\ee
with respect to the tidal problem, the expression above is analogous 
to the equilibrium tide.  The term $\propto \ddot{X}$ is small in comparison to the 
$X^{(0)}$ and does not contribute to energy transfer.  To next order in 
forcing frequency 
$\sigma=2\pi/P$ the contribution to
$X$ is due dissipation and is given by 
\be
X^{(1)}=-\,\gamma\,\dot{X}^{(0)}=-\,\tau\,\dot{F}
\ee
which allows us to write
\begin{eqnarray}
X & = & \frac{1}{\sqrt{2}}\frac{F_0}{\omega^2}\,e^{i\sigma\, t-i\delta}+{\rm c.c.}
\end{eqnarray}
where the time lag is given by 
\be
\tau=\gamma/\omega^2
\ee
and the phase lag $\delta=\sigma\,\tau$.

For a damped, driven harmonic oscillator $Q$ is defined as
\be
\frac{1}{Q}=\frac{-\oint_{\rm cyc} dt\,{dE}/{dt}}{2\pi\,E_0}=\frac{-\Delta E_{\rm cyc}}{2\pi\,E_0}
\ee
where $\Delta E_{\rm cyc}$ is the energy dissipated per forcing cycle and is given by 
\be
-\Delta E_{\rm cyc} =\oint_{\rm cyc}dt\,\dot{X}F \simeq 2\pi
\frac{\abs{F_0}^2}{\omega^2}
\,\delta
\ee
and the peak energy $E_0$, is the energy stored in the interaction 
\be
E_0= \left<X\,F\right> \equiv \frac{1}{P}\oint_{\rm cyc}dtX\,F=\frac{\abs{F_0}^2}{\omega^2}
\ee
and therefore, 
\be
\frac{1}{Q}=-\frac{\oint_{\rm cyc} dt\dot{X}\,F }{2\pi\left<X\,F \right>}= \delta=\sigma\,\tau.
\label{e: Q_osc}
\ee

The oscillator $Q$ is used as a proxy for the phase lag
of the forced response.  Its definition requires that the time-dependance 
of the forcing be described by a single Fourier harmonic.  Given the forcing
frequency $\sigma$, a constant lag time $\tau$ is also required to 
define $Q$.  

In general, the forced damped harmonic oscillator is not driven by a
a single Fourier harmonic and the time-dependance can be 
completely arbitrary.  The more general solution 
in the non-resonant weak friction limit may be written as
\begin{eqnarray}
X(t)&=&X^{(0)}+X^{(1)}=X^{(0)}-\tau\dot X^{(0)}\nonumber\\
&\simeq&X^{(0)}(t-\tau)
\end{eqnarray}
in the limit that $\tau$ is small and thus
the response $X(t)$ follows the forcing $F(t)$ by some
time lag $\tau$. Note that the lag time $\tau$ parameterizes the level 
of dissipation for a damped harmonic oscillator, for arbitrary 
forcing.  The oscillator $Q$, which is equivalent to the phase
lag $\delta$, characterizes the level of dissipation for the specialized
case of sinusoidal forcing.  Unlike $\omega$ and $\tau$, 
$Q$ is not an intrinsic property of the oscillator as it depends on a
particular property of the external driving force i.e., the forcing frequency $\sigma$.

As previously mentioned, Socrates \& Katz (2012) show that for the tidal problem, a constant lag time $\tau$
is an intrinsic property of the forced body, given the physical requirements
listed in \S\ref{ss: basis}, assuming the that fluid response of the forced
body obeys the following equation of motion
\be
\ddot{\bxi}+{\bf C}\cdot\bxi+{\bf D}\cdot\bxi=-\grad U_T
\label{e: EOM}
\ee
where $\bxi$ is the Lagrangian displacement and $U_T$ is the tidal 
potential.  The operators ${\bf C}$ and ${\bf D}$ can be thought 
of as a generalized ``spring" coefficient and damping rate, respectively.

The tidal potential can be decomposed into multipoles of spherical 
harmonic degree $\ell$.  The lag time $\tau_{\ell}$ of the corresponding
tidal multipole response $q_{\ell m}$ is given by
\be
\tau_{\ell}=\frac{\int d^3x\,\rho\, {\bf \Xi}^*_{\ell m}\cdot{\bf C^{-1}} \cdot{\bf D}\cdot{\bf C}^{-1}
\cdot{\bf \Xi}_{\ell m}}{\int d^3x\,\rho\,{\bf\Xi}^*_{\ell m}\cdot{\bf C}^{-1}
\cdot{\bf \Xi}_{\ell m}}
\ee
where ${\bf\Xi}_{\ell m}({\bf x})=\grad r^{\ell}Y_{\ell m}({\bf \Omega})$
describes the spatial variation of the tidal field about the co-ordinates, 
${\bf x}=\left(r,{\bf\Omega}\right)$, of the forced body.  Both ${\bf C}$
and ${\bf D}$ are time-independent
differential operators that depend, like the fluid density $\rho$, on the 
internal structure of the forced body.  It then follows that, as in the case of the
non-resonant weakly damped forced harmonic oscillator, the lag time is
an intrinsic property of a non-resonant weakly damped tidally forced 
body.

Socrates \& Katz (2012) also show that there is a constant
time lag $\tau_{\ell}$ for each spherical harmonic degree 
$\ell$ of the forced response.  By adopting the approximation 
that the perturber is distant such that only the $\ell=2$
terms are kept, they are then able to reproduce the basic
assumptions of Hut (1981).

Even in the event that the time dependence of the tidal 
forcing is described by a single Fourier harmonic, it 
is not clear why, and under what conditions, 
the level of dissipation of the tidal problem 
can be parameterized by a quality factor $Q$.  As previously mentioned, 
the tidal potential is the sum of the tidal harmonics, each of which 
may be a sum of Fourier harmonics.  
Furthermore, and on a more 
basic level, the left hand side of eq. \ref{e: EOM} is not, in general, 
equivalent to the left hand of eq. \ref{e: harm_osc}.  In other words, the tidal 
problem is not equivalent, in general, to a forced damped harmonic oscillator.

Nevertheless, in the limiting cases where the constant time 
lag model is valid and when there is a single forcing frequency, $\sigma$, the
tidal quality factor $Q$ is given by the ratio of the energy dissipated during 
the tidal forcing cycle and the average tidal interaction energy of the oscillating 
components give by (see Appendix)
\be
\frac{1}{Q}
 =-\frac{\sum_{\ell=2, m}\oint_{\rm cyc}\dot{q}^*_{\ell m}\Psi_{\ell m} }{\sum_{{\rm osc},
 \ell=2, m} 2\pi
 \left<q^*_{\ell m}\Psi_{\ell m}\right>} = \delta =\sigma \tau.
\ee
here $q_{\ell m}$ is the induced multipolar moment arising from the presence 
of the (normalized) tidal potential $\Psi_{\ell m}$ and the $\sum_{\rm osc}$
in denominator enforces the inclusion of only the oscillating components
of $q_{\ell m}$ and $\Psi_{\ell m}$.  
Note the correspondence between the expression above and the expression 
for the oscillator $Q$ given by eq. \ref{e: Q_osc}.  In the Appendix 
we show that, under several limiting conditions, the tidal $Q$ as defined
above is equivalent to a single phase lag.

\subsubsection{e=0; synchronization}

Both the $m=\pm 2$ components of the tidal potential oscillate sinusoidally
at the same frequency $\sigma=2\abs{\Omega-n}$.  The $m=0$ perturbation 
is time-independent in this orbital configuration and therefore, does not 
contribute to secular evolution. 
It follows that the tidal $Q$ is well defined as discussed above.
As we state in the Appendix, the peak energy $E_0$
is due to equal contributions of both components.  Furthermore, 
since the lag time $\tau$ is independent of $m$, the forced response corresponding
to each component of the tidal forcing possess the same phase lag (and angle)
as well and therefore, the same $Q$.

By using $\Delta E$ from eq. \ref{e: Delta_E} and $E_0$ from eq. \ref{e: E_0_synch}
we may write the familiar relation
\be
\frac{1}{Q_{_{\rm synch}}}=2\,\abs{\Omega-n}\tau.
\label{e: Q_synch}
\ee
In the expression above we made use of the fact the orbital period and the period 
of the forcing cycle differ by a ratio of $2\abs{\Omega-n}/n$ in order to relate
$\Delta E_{\rm cyc}$ to $\Delta E$ of eq. \ref{e: Delta_E}.

\subsubsection{e $\ll$ 1; circularization}

Again, the tidal $Q$ is well defined for this orbital configuration in part because
the tidal forcing is described by a single forcing frequency $\sigma=n$.
The forcing is composed of three distinct Fourier harmonics that originate from 
the $m=0$ and $m=\pm 2$ components of the tidal potential.  As we state in the Appendix, the peak energy $E_0$
is the sum of all three components, where the contributions from the $m=0$ and
$m=\pm 2$ components are not equal.  Furthermore, 
since the lag time $\tau$ is independent of $m$, the forced response corresponding
to each component of the tidal forcing possesses the same phase lag (and angle)
as well and therefore, the same $Q$.  Consequently, the level of dissipation 
can be parameterized by a single $Q$ as well.

By using $\Delta E$ from eq. \ref{e: Delta_E} in the limit that $\Omega=\Omega_{\rm ps}$ (cf. 
eq. A 36 of Hut (1981)) and 
along with the expression for the peak energy $E_0$ from eq. \ref{e: E_0_circ} we may
write
\be
\frac{1}{Q_{_{\rm circ}}}=n\,\tau.
\label{e: Q_circ}
\ee
Here, $\Delta E$ of eq. \ref{e: Delta_E} is equal to $\Delta E_{\rm cyc}$ since 
the forcing frequency is equal to the mean motion $n$.

\subsubsection{e $\approx 1$; high-e migration}

In this limit, $Q$ is not well defined i.e., there is neither a well-defined forcing cycle
in order to compute an $\oint_{\rm cyc}$-type integral nor a well-defined phase
lag since the forcing is not sinusoidal.  Nevertheless, it is possible to define a 
forcing cycle that is {\it approximately periodic} in the limit that $e\rightarrow 1$ since
tidal potential and thus $\dot E$ vanishes at apoastron.  In order to define a $Q$ for 
high-$e$ migration a peak energy $E_0$ is needed, which we set to the peak interaction given by 
\be
E_0=\frac{k_{L}\,GmR^5}{r^6_p}
\ee
where $r_p=a(1-e)$ is the periastron distance.  Now we may define
a tidal $Q$ during high$-e$ migration  
\be
\frac{1}{Q_{_{\rm HEM}}}\equiv-\frac{\Delta E}{2\pi\,E_0}=\frac{3861}{4096}\,n_{\rm F}\,\tau.
\label{e: Q_high_e}
\ee
Interestingly, $Q_{_{\rm HEM}}$ as defined above remains approximately a constant during high-$e$
migration.  Take for example, a case where the initial value of eccentricity is
$e=0.999$ and tidal dissipation shrinks the orbit to a value of $e=0.99$.  Though the 
orbital period changes by a factor of $\sim 30$, the value of $Q_{_{\rm HEM}}$ changes by 
less than 1\%.  The shape of the orbit
near periastron -- where all of the energy transfer takes place -- remains approximately
constant during evolution (e.g., Socrates et al. 2012).  It follows that $\Delta E_{\rm cyc}=\Delta E$ per orbit
remains approximately a constant as does the peak energy $E_0$ and therefore, 
so should $Q_{_{\rm HEM}}$.

\subsection{other parameterizations}

Within the literature, there are other parameterizations 
of the constant lag time for various theoretically-based motivations.  
Expressions relating them to the lag time $\tau$ are given below.  

Hut (1981) defines a 
characteristic time scale for dissipation
\be
T\equiv \frac{R^3}{G\,M\,\tau}
\ee   

Eggleton et al. (1998; see also Fabrycky \& Tremaine 2007) use the 
viscous time $t_V$  
\be
t_V=3\left(1+1/k_{L}\right)T=3\left(1+1/k_{L}\right)\frac{R^3}{G\,M\,\tau}
\ee
and the coefficient $\sigma_V$ incorporated by Hansen (2010; 2012)
\be
\sigma_V=\frac{2}{3}k_{L}\tau\,G\,R^{-5}.
\ee

\section{Application to gas giant planets}
\label{s: calibration}

\begin{center}
\begin{deluxetable}{cccc}
\tabletypesize{\scriptsize}
\tablecaption{Inferred values of various tidal dissipation parameters for Jupiter-like
planets
\tablenotemark{a}
\label{t: parameters}}
\tablewidth{0pt}
\tablehead{\colhead{}
 & 
\colhead{ Jupiter-Io}& \colhead{hot Jupiter }&\colhead{high$-e$} \\
\colhead{parameters} & 
\colhead{ interaction}& \colhead{circularization }&\colhead{ migration\tablenotemark{b} }  }
\startdata
$\tau$ (s)& 0.062  & $0.25$ & $0.66$ 
\\
$Q$& $5.9\times 10^4$& $2.8 \times 10^5$ & $1.1\times 10^5$
\\
$t_D$ (Gyr)& $9.5\times 10^5$ & 20 & $7.5$ 
\\
$T$ (hrs)& $1.2\times 10^4$ & $3.0\times 10^3$& $1.1 \times 10^3$ 
\\
$t_V$ (hrs)& $1.3\times 10^5$ & $3.3\times 10^4$& $1.2 \times 10^4$ 
\\
$\sigma_V$\tablenotemark{c}($10^{-59}$)& $6.3$ & $25$& $67$ 
\\
\enddata 
\tablenotetext{a}{We take the Love number $k_L=0.38$, $R_p=6.99\times 10^9{\rm cm}$, $M_p=1.9\times
10^{30}$g, $m_{_{\rm Io}}=8.93\times 10^{25}$g, 
the mass of the stellar companion = $M_{\odot}$ and
the final circularization distance $a_{_{\rm F}}=0.06$ AU. }
\tablenotetext{b}{For high$-e$ migration the tidal $Q$ is not well-defined
and we use $Q_{\rm HEM}$ as defined by eq. \ref{e: Q_high_e}.}
\tablenotetext{c}{The 
parameter $\sigma_V$ has dimensions of $\left({\rm g\,cm^2\,s}\right)^{-1}$.}
\end{deluxetable}
\end{center}

In what follows, we determine the values of $\tau, t_V,$ and $\sigma_V$
for Jupiter, based on its interaction with Io.  We then compare these values 
with the required values for circularization and high$-e$ migration of
Jupiter analogues ($M=M_J,\,R=R_J$ and same structure).  In particular, 
we estimate the value of $Q$ and $t_D$ for these cases.

\subsection{Jupiter-Io interaction}
\label{ss: J-Io}

Jupiter's rotation is super-synchronous with respect to its orbit 
with Io ($\Omega_J > n_{_{\rm Io}}$).  In order
to obtain a timescale for the evolution of Io's orbit, use eq. \ref{e: adot_synch} 
and integrate from $a=0$ to $a=a_{\rm Io}$ in order to write
\be
\frac{a}{\dot{a}}\vert_{_{a=a_{\rm Io}}}=\frac{13}{2\,\Lambda}t_{\odot}
\ee
where $t_{\odot}=4.5$ Gyr is the age of the Sun and
\be
\Lambda=\frac{\sum_{i={\rm Io}}^{
{\rm Ganymede}} L_i}{L_{\rm Io}}\approx 4.3
\ee
where $L_i$ is the orbital angular momentum if the $i^{\rm th}$ Galilean moon.
In obtaining an upper limit for dissipation, the inclusion of the factor $\Lambda$
conservatively assumes that the current resonant configuration of the inner three Galilean 
moons is preserved during outward migration.  From the form of 
$t_D$ given by eq. \ref{e: t_D}, we obtain a limit of the lag time $\tau_{_{\rm J-Io}}$
for the Jupiter-Io interaction
\be
\tau_{_{\rm J-Io}}\leq 0.062\,{\rm s}
\label{e: tau_JIo}
\ee    
where we have set $k_L=0.38$.  By using eq. \ref{e: Q_synch}
we may determine the lower-limit for $Q_{_{\rm J-Io}}$, the 
tidal $Q$ for the Jupiter-Io interaction
\be
Q_{_{\rm J-Io}}\geq 5.9\times 10^4
\ee
as found by other authors
(Goldreich \& Soter 1966; Yoder \& Peale 1981; Leconte et al. 2010).

\subsection{circularization of hot Jupiters}

Nearly all hot Jupiters with orbital periods $P\leq 5$ days and semi-major axes
$a\leq 0.06$ AU have circular orbits, while the values of $e$ belong to a broad
distribution for gas giants at larger separation.  A common interpretation is 
that planets that are relatively close-in circularize quickly due to tidal 
dissipation's strong dependance on orbital separation.  Consequently, it
is possible to obtain a separate constraint on the dissipation parameters 
$\tau, Q,$ etc., by presuming that tidal dissipation is responsible for 
circularizing Jupiter-like planets with $P\leq 5$ days within a
characteristic age $t_0=10$ Gyrs of these systems.  

If we set $t^{-1}_0=\dot{e}/e$, given by the right hand side of eq. \ref{e: edot_circ}
then we can write an expression for $t_D$
\be
t_D=2\,t_0.
\ee
With the help of eq. \ref{e: t_D} and 
by setting $k_L=0.38$ and $a_{\rm F}=0.06$ AU we arrive at a constraint for $\tau$
\be
\tau\geq 0.25\,{\rm s}
\ee
for a Jupiter-analogue in orbit around a Sun-like star.  In the above 
constraint, we chose $k_2=0.38$ and $a_{\rm F}=0.06$ AU. 
Given this value of $\tau$, eq. \ref{e: Q_circ} enables us to determine $Q$
\be
Q_{_{\rm circ}}\leq 2.8\times 10^5.
\ee

\subsection{high-e migration of hot Jupiters}
\label{ss: high-e}

A leading theory for the formation of
hot Jupiters is high-$e$ migration (see Socrates et al. 2012 
for a review).  In particular Fabrycky \& Tremaine (2007)
predicted that, as a result of high-$e$ migration, the orbital
angular momentum of hot Jupiters should, in general, be mis-aligned 
with the spin axis of the stellar host, which was later confirmed
observationally by Winn et al. (2010).  

The migration time $t_m$ between an initial eccentricity $e_i$
and some final value $e_0$ can be found by numerically integrating eq. \ref{e: adot_high_e} as
depicted in Figure 1 of Socrates et al. (2011).  For example, a Jupiter analogue 
starting at an orbital period $P\approx 12$ yrs
that circularizes to a final distance $a_{\rm F}\approx 0.06$ AU with $e_0\approx 0.1$
migrates in a time $t_m$ given by
\be
t_{\rm m}=t(e_{\rm i}=0.994)-t(e_{0}=0.1)\approx \frac{4}{3}t_{D}.
\ee
For a fixed value of $J$ or $a_{\rm F}$,
the migration time $t_m$ does not vary considerably if the initial value of the orbital 
period varies at the order unity level.
For high$-e$ migration to operate the above expression requires that $t_D$ satisfies
\be
t_D\lesssim \frac{3}{4}\,t_0
\ee
or the lag time obeys
\be
\tau\gtrsim 0.66\,{\rm s}
\label{e: tau_HEM}
\ee
for a Jupiter analogue.

With our definition given by eq. \ref{e: Q_high_e}
along with the value of $\tau$ above to place a limit on $Q$ during high-$e$
migration
\be
Q_{_{\rm HEM}}\lesssim1.1\times 10^5.
\ee
Note that $t_m$ and consequently, $\tau$ and $Q_{_{\rm HEM}}$
are rather insensitive to $e_i$.

\subsection{discussion}

If indeed Jupiter analogues with $P_{\rm F}=5$ days successfully undergo high$-e$
migration, then according to the results of 
\S\S\ref{ss: J-Io} and \ref{ss: high-e}, {\it there is at least an order of magnitude
disagreement between the two requirements}.  In other words, the time $\tau$ required
for high$-e$ migration is ten times larger than the allowed upper-limit for $\tau$
of Jupiter.  For example, 
if Jupiter's lag time $\tau_{_{\rm J-Io}}=0.66$ s, as required for high-$e$ migration, then 
its tidal quality factor would decrease by an order of magnitude such that
$Q_{_{\rm J-Io}}\lesssim 6\times 10^3$.

The tension between two requirements for $\tau$ is obscured when tidal dissipation 
is quantified by $Q$.  At face value, the requirements for $Q$ from circularization and high$-e$ migration indicate that less energy is dissipated per cycle than that inferred from 
the Jupiter-Io calibration.  However, as detailed in \S\ref{ss: Q}, the tidal $Q$, 
unlike the lag time $\tau$, is not an intrinsic property of the forced body in the 
equilibrium theory.  Consequently, if the determination of $\tau$ from the 
Jupiter-Io interaction is correct, then Table \ref{t: parameters} reveals that a
Jupiter analogue would not be able to migrate from its initial position 
to $P_{\rm F}\approx 5$ days within the Hubble time.

Note that the requirement on $\tau$ from 
tidal circularization of hot Jupiters is marginally consistent with the 
value inferred from the Jupiter-Io interaction, which is roughly in agreement
with several previous studies (e.g., Leconte et al. 2010; Hansen 2010 \& 2012).

\section{previous work on high-eccentricity migration}  
\label{s: previous}

The fact that the lag time $\tau$ necessary to migrate a Jupitier analogue to 
a final circularization period $P_{\rm F}\approx 5$ days is at least 10 times
larger than $\tau_{\rm J-Io}$ does not present an enormous hurdle for 
high-$e$ migration scenarios.  Due to the strong dependence of $t_{D}$ on 
the final circularization radius $a_{\rm F}$ and planet radius $R_p$, it is possible that
a factor of 10 may be recovered in order to make the migration time $t_{\rm m}$ 
consistent with $t_0\lesssim 10$ Gyr.

However, there is a general understanding in the literature that
for modest values of tidal dissipation inferred from the Jupiter-Io interaction 
($Q_{_{\rm J-Io}}\approx {\rm a \, few\times 10^5}$), the migration 
time of a Jupiter analogue with $P_{\rm F}\approx 5$ days is 
of order $\sim 300$ Myrs (Wu \& Murray 2003; Fabrycky \& Tremaine 2007)
in stark contrast with our results of the last section.

\subsection{Wu \& Murray 2003}
\label{ss: WM}

Wu \& Murray (2003; hereafter WM) produced the first calculations
of high-$e$ migration for hot Jupiters, with particular emphasis on 
HD 80606b.\footnote{In a planetary context, high$-e$ migration was
first utilized to explain the misalignment of Triton's orbit with 
the spin of Neptune by Goldreich et al. (1989).  Those authors do not
provide justification for their prescription for tidal dissipation.  In the high$-e$ limit, the behavior of their evolutionary equations is, 
correctly, identical to that of Hut (1981).  }    Their evolutionary equations are taken from 
Eggleton et al. (1998).  In this model, tidal dissipation is ultimately 
modeled after the orbit-averaged constant time lag model of Hut (1981).  In order
to determine the dissipation rate, a constant time lag $\tau$ must
be specified.  

However, inspection of their eq. (A9) reveals that the value of $\tau$ in their
calculations is
\be
\tau{_{\rm WM}}=\frac{1}{n\,Q_0}=192.8 \left(\frac{3\times 10^5}{Q_0}\right)
\left(\frac{P}{12\,{\rm yrs}}\right)\,{\rm s}
\label{e: WuMu_tau}
\ee
where $n$ is the mean motion and they set $Q_0=3\times 10^5$.
Note that $\tau_{_{\rm WM}}$ is not a constant during evolution and
may be $\sim 3000$ times larger than upper limit for $\tau_{_{\rm J-Io}}$
given by Table \ref{t: parameters}.  This particular
deformation of the constant $\tau$ model of equilibrium tides results in 
an infinite amount of energy dissipated per orbit in the limit that $e\rightarrow 1$
and $P\rightarrow\infty$ for a fixed periastron.  Finally, in this $e\rightarrow 1$ limit, the tidal 
$Q_{_{\rm HEM}}=-2\pi E_0/\Delta E\rightarrow 0$.
 
Similar errors
can be found in related works (Wu et al. 2007; Wu \& Lithwick 2010).
For example, consider eq. 5 of Wu et al. (2007), which they state is
equivalent to eq. 10 of Hut (1981), in fact reveals that
\be
\frac{d\,{\rm ln}\,e}{dt}\Big|_{\rm Wu}\sim \frac{d\,{\rm ln}\,e}{dt}\Big|_{\rm Hut}\times \frac{P}{P_{\rm F}}.
\ee
Since the orbit is pseudo synchronized during high$-e$ migration,
$P_{\rm F}$, $J$ ad $a_{\rm F}$ are all fixed during evolution.  Therefore, 
for a orbit starting with $P=12$ yrs migrating on a track of 
angular momentum $J$ corresponding to $P_{\rm F}=5$ days, the lhs 
of the expression above exceeds the rhs by a factor $\sim P/P_{\rm F}\sim 10^3$.

\subsection{Fabrycky \& Tremaine 2007}
\label{ss: FT}

Like WM, Fabrycky \& Tremaine (2007; hereafter FT) employ
the evolutionary equations of Eggleton  et al. (1998), where the prescription
for tidal dissipation is based upon Hut (1981).  Unlike WM however, 
they did not alter the orbit evolutionary equations.  They qualitatively 
reproduce the presumed migration history of HD 80606b calculated
by WM.  In order to do so, they used a viscous time $t_V$ of
\be
t_{V_{\rm FT}}=0.001\,{\rm yr}=8.8\,{\rm hrs}.
\label{e: t_V_FT}
\ee
The qualitative agreement between WM and FT seems to strengthen 
the case for high-$e$ migration with respect to the history of HD 
80606b.  When computing the distribution 
of the inclined hot Jupiters, FT use a value for $t_V$ of
\be
t_{V_{\rm FT}}=0.01\,{\rm yr}=88\,{\rm hrs}.
\ee

However, while FT used the correct equations of motion their value for
$t_V$ departs from the value determined from the Jupiter-Io interaction given in  
Table \ref{t: parameters} 
\be
t_{V_{\rm J-Io}}=1.3\times 10^5 \,{\rm hrs},
\label{e: t_V_J-Io}
\ee
by four orders of magnitude.  Furthermore, note that the value FT 
use for $t_V$ in both cases leads to a value of $\abs{\Delta E/E_0}\approx 0.1$ during 
high-$e$ migration.

In addition, FT write an expression in their equation A10 for the tidal 
$Q$ that follows
\be
Q\propto P
\ee
which as previously discussed at length, is incorrect within the context
of high-$e$ migration.

\subsection{discussion of previous work}

The mis-calibrations highlighted above demonstrate the danger of using
$Q$ when parameterizing the strength of tidal dissipation.  For the case of the
synchronization tide in the Jupiter-Io system, there is little quantitative difference
between using $Q$ and $\tau$ since the forcing period of Jupiter is approximately 
constant through the evolution of Io's orbit.  Consequently, in the equations of 
motion, replacing lag time -- or phase lag -- in the equations of motion can 
be done without loss of accuracy e.g.,  as in Goldreich \& Soter (1966). 
Nevertheless, due to the manner in which $Q$ is defined, its placement
within equation of motion (cf. \S II of Goldreich \& Soter 1966) can
be extremely confusing or incorrect altogether.

The intuition derived
from Goldreich \& Soter (1966) for the synchronization tide of Jupiter cannot
be applied to the more general tidal problem, where the orbit is eccentric.
In the case of the Jupiter-Io interaction, a 
phase lag and physical lag angle are equal to one another, both of which 
are equivalent to the energy dissipated over a well-defined periodic sinusoidal 
forcing cycle.  There is, in general, no well-defined forcing cycle in the tidal 
problem and therefore, no single phase lag.  Furthermore, for the example 
of a pseudo synchronized eccentric orbit, the lag angle is a function of 
orbital phase, where its value at e.g., apoastron and periastron are not equal.

Despite previous mis-calibrations in the level of tidal dissipation in migrating super-eccentric
gas giants, as previously mentioned, Table \ref{t: parameters} indicates
that a modest increase by a factor of $\sim 10$ in the relative strength of dissipation 
allows for Jupiter analogues to migrate  from a period of 12 years to 5 days within the age
of the Universe.  This is a direct consequence of the fact that tidal dissipation 
is an extremely strong function of orbital angular momentum $J$, periastron $r_p$
or $a_{_{\rm F}}$.  For example, $a_{_{\rm F}}=0.071$ AU in Figure 1 of FT, rather than 
HD 80606b's value for $a_{_{\rm F}}=0.061$ AU, dissipation is also intrinsically weaker 
since they chose $M_p=7.8\,M_J$, migration was completed in $\sim 3-4$ Gyrs, rather than $t_0=10$
Gyrs, at long orbital periods the fractional amount of time migrating was small since
the planet underwent Kozai oscillations and therefore, spent a significant
fraction of its history at high$-J$.  By accounting for all these
effects, well over two of the orders that separate eqs. \ref{e: t_V_FT} and \ref{e: t_V_J-Io}
may be erased .  In what follows, we discuss some of the consequences for 
studies of gas giants that result from the requirement of
modestly enhanced dissipation for high-$e$ migration.

\section{Some Implications for studies of gas giants}\label{s: implications}

\subsection{tidal theory}

It was realized early on that turbulent convection is severely inadequate in providing
the required amount of dissipation in order to account for Jupiter's tidal $Q$
(Goldreich \& Nicholson 1977).  Since then, the leading candidate for the source
of tidal dissipation has been the resonant excitation of low frequency, nearly incompressible,
waves that are primarily restored by the Coriolis force.
This topic is technically challenging and continues to be
in a state of development (Ogilvie \& Lin 2004; Arras 2004; Wu 2005a,b; Goodman \& Lackner 2009;
Ogilvie 2009).  

The results and relations of equilibrium tide theory cannot be directly 
compared with predictions of dynamical tidal theories.
Nevetheless, the value of $\Delta E$ required for circularization 
for a given $P_{\rm F}$ within the Hubble time is largely independent of 
any particular tidal model.  Goodman \& Lackner (2009) write an 
expression for a tidal $Q=-\Delta E/2\pi E_0$ during hot Jupiter circularization that results from 
the inertial wave model.  In their expression for $Q$, their peak energy $E_0$
is the peak energy in the fluctuating component of the interaction energy, appropriate 
for circularization given by our eq. \ref{e: E_0_circ} in the Appendix.
 
Their reported  value for $Q$ is $Q\gtrsim 3\times 10^7$
(e.g., Goodman \& Lackner 2009) for Jupiter analogues (core radius $R_c\lesssim 0.2 R_J$) with $P_{\rm F}=5$ days.  
Table \ref{t: parameters} indicates
that the required $Q$
for circularization of hot Jupiters with $P_{\rm F}\approx 5$ days is $Q\leq 3\times 10^5$.
Thus, calculations of tidal circularization of gas giant planets arising from 
the inertial wave mechanism fall short by a factor of 100 or so.  High-$e$ migration 
will then be almost certainly more difficult to accomplish by the inertial 
wave mechanism.  However, a calculation of the forced response that leads to the
inertial wave mechanism for a highly eccentric orbit has yet to be performed, so 
a quantitative comparison with results from equilibrium tidal theory are therefore, not yet possible.

A clear difference between the properties of the tidal potential felt by 
gas giants undergoing high-$e$ migration and Jupiter is the amplitude 
of the tidal potential.  The ratio is given by $U_{_{\rm Io}}/U_{\star} \approx 10^{-3}$.  If 
indeed super-eccentric migrating Jupiters
experience enhanced tidal dissipation during migration, then it may be due to
non-linear effects.

\subsection{the inflated radii problem}

The hot Jupiters are inflated in that, unlike Jupiter, their radii
are, in general, significantly larger than the zero temperature solution.
There are several solutions to this problem, many of which depend upon 
tidal dissipation in one way or the other.

There is a class of mechanisms that can be thought of ``delayed" contraction
mechanisms.  Here, the planet is brought into its close orbit in such 
a way that its degeneracy is lifted (Burrows et al. 2007; Leconte et al. 2010; 
Ibgui \& Burrows 2009; Batygin \& Stevenson 2010; see also Wu \& Lithwick 2011).  
One candidate is tidal dissipation due to circularization during high-$e$ migration.
Table \ref{t: parameters} indicates that in order for these mechanisms to work, 
tidal dissipation must be much stronger for super-eccentric migrating Jupiters
in comparison to that inferred from the Jupiter-Io interaction.

Thermal tidal torques can render the hot Jupiters in a persistent state of asynchronous
spin.  The resulting ongoing dissipation of the gravitational tide, presumably at great depth,
allows for a steady source of thermal energy required to inflate the radii significantly
above the zero temperature solution (Arras \& Socrates 2009a, 2009b; 2010).  In order
for this mechanism to work, the out-of-phase quadrupole induced by thermal forcing
near the optical photosphere must be equal to the out-of-phase quadrupole induced 
by dissipation of the gravitational tide.  Such an equilibrium can only be realized if the 
relative level of tidal dissipation in the hot Jupiters is comparable to that inferred from 
the Jupiter-Io interaction.  Therefore, if high-$e$ migration is indeed responsible for 
delivering gas giants from initial orbital periods of $\approx 12$ yrs to a  
final circularized state of $\approx 5$ days, then  thermal tidal torques
cannot compete against the gravitational tidal torques.  All of this, however, 
is under the assumption that the lag time for the synchronization and circularization
tide are equal, which may not be the case in reality.

\section{Summary}\label{s: summary}

If high-$e$ migration leads to the production of hot Jupiters 
with final circularized orbital periods $P_{\rm F}\approx 5$ days, tidal dissipation in these
extrasolar gas giants must be at least 10 times stronger than that
inferred from the Jupiter-Io interaction.  While this alone cannot rule out
high$-e$ migration, it is in severe conflict with several previous studies of high-$e$ migration. 
We showed that the discrepancy between our analysis and these studies arises from 
the fact that they employed dissipation strengths which are 
$10^3-10^4$ larger than the inferred upper limit of the Jupiter-Io interaction.

While current theories of tidal excitation and dissipation in gas giant
planets can account for the inferred level of dissipation in the Jupiter-Io 
interaction (Ogilvie \& Lin 2004) they cannot currently account for the amount 
of dissipation required for the observed circularization of hot Jupiters. 
If high$-e$ migration is in fact responsible for hot Jupiter migration, the discrepancy between 
the level of computed dissipation in the inertial wave mechanism and that required
to account for migration is even larger.  If, for example, thermal emission 
from migrating Jupiters in orbit about late-type stellar hosts
are observed by on-going and future direct-imaging campaigns (Dong et al. 2012),
then the tidal dissipation time $t_D$ cannot be much larger than $t_D\lesssim$ Gyr.

In the Appendix we point out that the the tidal quality factor is 
well-defined and can be related to a phase lag in 
only in two limiting cases:  asynchronous and circular orbits or 
pseudo-synchronous orbits with small eccentricity.  We find the relation 
between the tidal quality factor $Q$ and the lag time $\tau$ for these two limiting
cases previously mentioned and additionally during high-$e$ migration.

Given the results presented here, there is now
tension between various
sub-fields of gas giant studies: the theory of tidal friction in gas giant planets, 
few-body dynamics of exoplanet systems, theory of inflated radii, the
evolutionary scenario outlined by Goldreich \& Soter (1966) for the
Jupiter-Io interaction and even perhaps planet formation.
Either the standard thinking in one (or more) of these categories requires
revision or there are extra ingredients at work in forming (stellar spin-orbit)
mis-aligned hot Jupiters.  The production
of hot Jupiters is not a solved problem, though high-e migration is still viable.

\acknowledgements{We thank Phil Arras and Scott Tremaine 
for many helpful discussions.  We thank Daniel Fabrycky and Yanqin Wu for comments
and suggestions on the manuscript.  We also that Jeremy Goodman, Dong Lai, Yoram Lithwick, John Papaloizou, Eliot Quataert, and Matias Zaldarriagga for stimulating discussions.  Work by SD was performed under contract with the California Institute
of Technology (Caltech) funded by NASA through the Sagan Fellowship
Program. B.K was
supported by NASA through Einstein Postdoctoral Fellowship awarded by
the Chandra X-ray Center, which is operated by the Smithsonian
Astrophysical Observatory for NASA under contract NAS8-03060. AS acknowledges
support from a John N. Bahcall Fellowship awarded by the Institute for Advanced
Study, Princeton.}

\appendix

\section{Interaction Potential, Love Number, Peak Energy and Energy Transfer Rate}

We review the mathematical machinery necessary to address 
the tidal problem.  We assume the forced object is perturbed by a source of gravity of
mass $m$
at distance $D(t)$ in a Keplerian orbit with true anomaly $\Phi(t)$ and that the strength 
of the tidal field is weak.  Therefore, only terms that are responsible
for linear perturbing forces and linear responses are kept.  The main purpose of what follows is to provide the necessary physical ingredients
that are required to connect the results of equilibrium tidal theories that utilize a constant time lag model for
tidal dissipation (Hut 1981) with those models that utilize a tidal quality 
factor $Q$ (cf. Goldreich \& Soter 1966).

\subsection{interaction potential}

To leading order in perturbation theory, the interaction potential is given by (cf. Newcomb 1962) 
\be
H_I=\int d^3x\,\rho\,\bxi\cdot{\grad} U_T
\ee
where $\int d^3x$ is taken over the volume of the forced body, $\rho({\bf x})$ is its static fluid density 
and $\bxi({\bf x},t)$ is the Lagrangian displacement field. The tidal potential $U_T$ is given by 
\begin{eqnarray}
U_T & = & -Gm\sum_{\ell m}\frac{4\pi}{2\ell+1}\left(\frac{r^{\ell}}{D^{\ell+1}}\right)
Y^*_{\ell m}\left(\pi/2,\Phi\right)Y_{\ell m}\left({\bf\Omega}\right)\nonumber\\
& = &-\sum_{\ell m}r^{\ell}Y_{\ell m}({\bf\Omega})\,\Psi_{\ell m}(t)
\end{eqnarray}
which serves to define $\Psi_{\ell m}(t)$.  Note that we we have specialized to the 
case where the perturber is located at co-latitude ${\theta}^{'}=\pi/2$.  Now, we may write
the interaction energy as
\be
H_I=-\sum_{\ell m}
q^*_{\ell m}\Psi_{\ell  m}
\label{e: H_I}
\ee
where $q_{\ell m}$ is the multipole moment (cf. Press \& Teukolsky 1977)
\be
q_{\ell m}\left(t\right)=\int d^3x\rho\,\bxi\cdot\grad r^{\ell}Y^*_{\ell m}\left({\bf\Omega} \right).
\ee
Note that for the $\ell=2$, the $\Psi_{\ell m}$'s carry 
dimensions of frequency squared.

\subsection{$k_{\ell}$ -- the Love number}

The Love number $k_{\ell}$ is defined as
\be
U^{\ell}_{\rm ind}(R)=k_{\ell}\,U^{\ell}_T(R)
\ee
where $U^{\ell}_{\rm ind}$ is the additional gravitational potential resulting for the 
induced tidal response of degree $\ell$ and  $R$ is the un-perturbed radius of the forced body.
The induced potential is given by (e.g., Jackson 1999)
\be
U_{\rm ind}=-G\sum_{\ell m}\frac{4\pi}{2\ell +1}q_{\ell m}r^{-\ell-1}Y_{\ell m}\left(\Omega\right)
\ee
and with this we can solve for each $q_{\ell m}$ in terms of $k_{\ell}$, $\Psi_{\ell m}$ and $R$
\be
q_{\ell m}=\Psi_{\ell m}/\varpi^2_{\ell}
\label{e: q_love}
\ee
where 
\be
\varpi^2_{\ell}\equiv\frac{4\pi \,G}{k_{\ell}\left(2\ell+1 \right)R^{2\ell+1}}.
\ee

\subsection{$E_0$ -- the peak energy}

The tidal $Q$ is a dimensionless number that parameterizes the relative
level of dissipation of the forced body over some interval in time.  
Given a tidal theory, a solution of the equations of motion allows for 
a determination of the energy dissipated $\Delta E$ over a forcing cycle.  
In order to construct the dimensionless parameter $Q$, which is equivalent
to a phase lag, a characteristic energy $E_0$ is needed.  Typically, 
$E_0$ is referred to as the ``peak energy.''  It is equal to the average energy
in the tidal interaction or the peak energy of the potential energy of the 
flow.  When computing $E_0$ only fluctuating components of the tidal 
potential are considered in order to maintain the standard relationship 
between $Q$ and the phase lag.  
In the literature, $E_0$ is defined in many ways and 
for the sake of clarity, we provide a definition for it.

Insert the expression for the induced multipoles in terms of $k$ and $\Psi_{\ell m}$
from eq. \ref{e: q_love} into the expression for $H_I$ and obtain
\be
H_I=-\sum_{\ell m}k_{\ell}\,\frac{2\ell +1}{4\pi}\frac{\abs{\Psi_{\ell m}}^2R^{2\ell+1}}{G}=-\sum_{\ell m}
\abs{\Psi_{\ell m}}^2/\varpi^2_{\ell}.
\ee
The interaction energy $H_I$ has contributions from both time-independent and 
fluctuating components. 
We define the peak energy $E_0$ with the following expression
\be
E_0\equiv-\left<H_I\right>=\frac{1}{P_{_{\rm cyc}}}\oint_{\rm cyc}dt\sum_{{\rm osc},\,\ell m}\abs{\Psi_{\ell m}^2}
/\varpi^2_{\ell}=\sum_{{\rm osc},\,\ell m}\abs{\Psi_{\ell m}^2}
/\varpi^2_{\ell}
\ee
where the integral is taken over the forcing cycle  and only terms resulting from 
the oscillating component of the tidal potential are kept.  

\subsubsection{synchronization:  $e=0$, $\Omega\neq \Omega_{\rm ps}$}

Tidal synchronization for circular orbits is due to forcing from the $\ell =2$ and $m=\pm 2$
portions of the tidal potential, which are responsible for its
time-varying component.  In this case, the oscillating component of the
tidal field can be thought of 
as a being due to a single Fourier harmonic with forcing frequency 
$\sigma=2\abs{n-\Omega}$.  The peak energy in the 
interaction $E_0$ is then given by  
\begin{eqnarray}
E_0 & = &-\left<H_I\right> =\sum_{\ell=2,m=\pm 2}
\abs{\Psi_{\ell m}}^2/\varpi^2_{\ell}\nonumber\\
E_0 & = & \frac{3\,k_{\ell}}{4}\frac{G\,m^2R^5}{a^6}.
\label{e: E_0_synch}
\end{eqnarray}
During synchronization, there is an $m=0$ tidal deformation as well.  However, 
since it is does not oscillate, it cannot contribute to energy transfer
and consequently, does not factor into the definition of $E_0$.

\subsubsection{circularization:  $e\ll1$, $\Omega=\Omega_{\rm ps}$}

For slightly eccentric orbits
with $e\ll 1$ in pseudo synchronous spin states with $\Omega\approx n$,
isolation of the oscillating terms of $U_T$ can be accomplished by 
a Fourier decomposition.  In the rotating frame, let 
\be
\left(\frac{a}{D}\right)^{\ell +1}e^{im\Phi}=\sum_{k=-\infty}^{\infty}
X_k^{\ell m}(e)e^{i\sigma_{km}t}
\ee
where the forcing frequency $\sigma_{km}=kn-m\Omega $ and $X^{\ell m}_k(e)$ are
the Hansen coefficients, which are given by 
\begin{eqnarray}
X_k^{\ell m}(e)&=&\frac{n}{2\pi}\int^{2\pi/n}_0 dt\,e^{im\Phi -i\sigma_{km}t}\left(\frac{a}{D}\right)^{\ell}
\nonumber\\
&\simeq &\delta_{km}+\frac{e}{2}\left[\left(\ell+1+2m\right)\delta_{k,m+1} +
\left(\ell+1-2m \right)\delta_{k,m-1}\right]+{\mathcal O}\left(e^2\right)
\end{eqnarray}
to leading order in eccentricity.  In the $e\ll1$, $\Omega=\Omega_{\rm ps}\approx n$
limit, the only terms that contribute to time-dependent driving are, for $\ell =2$, the 
$(m=0,k=1)$, $(m=2,k=2\pm 1)$ and the $(m=-2,k=-2\pm 1)$ terms.  Note that all 
of the oscillation frequencies are therefore given by $\sigma_{km}=\pm n$.

With decomposition into Hansen coefficients, the interaction energy $H_I$
and peak energy $E_0$ becomes
\begin{eqnarray}
E_0 & = &-\left<H_I\right> =   \sum_{{\rm osc},\,\ell m k}\abs{\Psi_{\ell mk}}^2/\varpi^2_{\ell}\nonumber\\
E_0 & = & \frac{21\,e^2\,k_2}{2}\frac{G\,m^2R^5}{a^6}
\label{e: E_0_circ}
\end{eqnarray}
where the $\Psi_{\ell m k}$'s are defined by the relation
\be
\Psi_{\ell m}=G\,m\,W_{\ell m}\frac{e^{i\,m\Phi}}{D^{\ell+1}}=\sum_{k}\frac{G\,m\,W_{\ell m}}{a^{\ell+1}}
X_k^{\ell m}(e)e^{i\sigma_{km}t}\equiv \sum_{k}\Psi_{\ell m k}\,e^{i\sigma_{km}t}
\ee
and the Fourier transform of the induced multipole moments $q_{\ell mk}$ are
\be
q_{\ell m}= \sum_{k}q_{\ell m k}\,e^{i\sigma_{km}t}
\ee
each of which satisfy 
\be
q_{\ell m k}=\Psi_{\ell m k}/\varpi^2_{\ell}.
\ee

\subsubsection{high-$e$ migration:  $e\rightarrow 1$, $\Omega=\Omega_{\rm ps}$}

For the case of high-$e$ migration, the time dependance of the tidal potential 
is far from that of a single Fourier harmonic.  Its magnitude is highly peaked
at periastron and falls of quickly with increasing 
distance.   We choose to define the peak energy as the entire interaction 
potential, since all of the $Y_{\ell m}$'s vary with time and therefore, contribute
to the potential energy available to perform work on the orbit.  We write
\begin{eqnarray}
E_0\equiv\max{\left[-H_I\right]}= k_2\frac{G\,m^2R^5}{r_p^6}
\label{e: E_0_high}
\end{eqnarray}
where $r_p=a(1-e)$ is the periastron distance.

\subsection{energy transfer rate}

The rate of energy transfer $\dot{E}$ between the orbit and internal degrees of freedom of
the forced body is given by
\be
\dot{E}=-\int d^3x\rho\,\dot{\bxi}\cdot\grad U_T.
\ee
The definition of $U_{T}$ in terms of the $\Psi_{\ell m}$'s may be employed in order to 
write
\be
\dot{E}=-\sum_{\ell m}\dot{q}^*_{\ell m}\Psi_{\ell m}.
\label{e: E_dot}
\ee

\section{conditions for a single Q from many oscillators and the tidal problem}

Even for the case of semi-diurnal forcing ($\ell =2;m=\pm 2$) during synchronization 
of a body in a circular orbit, the forcing may be thought of as originating from two driving terms.  It is not clear that the level of tidal dissipation in such a system can be summarized by a 
single $Q$, which itself is composed of a single phase lag.  In order to proceed, a
definition for $Q$ is required for a set of decoupled forced damped harmonic 
oscillators, each of which obey
\be
\ddot{X_n}+\omega^2_nX_n+\gamma_n\dot{X_n}=F_n(t)=\frac{F_{0,n}}{\sqrt{2}}\,e^{i\sigma_nt}+{\rm c.c.}
\ee
where $\omega_n$ is the fundamental frequency of each oscillator amplitude $X_n$, $\gamma_n$ is the corresponding damping rate and $\sigma_n$ is the driving frequency.  
Each of the $n$ oscillators has a $Q=Q_n$ given by
\be
\frac{1}{Q_n} = -\frac{\oint_{\rm cyc}dE_n/dt}{2\pi\,E_{0,n}}=-\frac{-\oint_{{\rm cyc}}{\dot {X_n}}{F_n}}{2\pi \left<X_n\,F_n\right>}=\sigma_n\,\tau_n =\delta_n
\ee
where $\tau_n=\gamma_n/\omega^2_n$ is the lag time and 
$\delta_n$ is the phase lag.   In the expression above $X_n(t)$ is the 
solution in the non-resonant weak friction limit where $X_n(t)=F_n(t-\tau_n)/\omega^2_n.$

Now, define $Q$ for the entire set of oscillators
\be
\frac{1}{Q}\equiv -\frac{\sum_n\oint_{\rm cyc} dE_n/dt  }{\sum_n 2\pi E_{0,n}}
=  -\frac{\sum_n\oint_{\rm cyc} dE_n/dt  }{\sum_n 2\pi \left<X_n\,F_n\right>} =\frac{\sum_n \delta_n\,\abs{F_{0,n}}^2 /\omega^2_n}{\sum_n \abs{F_{0,n}}^2 /\omega^2_n}
\label{e: Q_phase}
\ee
The expression above is well-defined only if there is a single forcing cycle and therefore, 
only a single forcing frequency.   
In the event that there is a single lag time for each oscillator, then $\delta_n=$
a constant.  It follows that
\be
\frac{1}{Q}=\delta_n=\delta.
\ee
Or in other words, a single $Q$ that is equivalent to a single phase lag
quantifies the level of dissipation of a set decoupled oscillators only
if their lag times are equal to one another.

To make contact with the tidal problem, we restrict to $\ell=2$ and 
write the tidal $Q$ as
\begin{eqnarray}
\frac{1}{Q}&=&-\frac{\oint_{\rm cyc}dE/dt}{2\pi\,E_0} =\frac{\oint_{\rm cyc}dE/dt}{2\pi\ \left<H_I\right>}  = 
\frac{\oint_{\rm cyc}dE/dt}{\sum_{\ell=2, m}2\pi\left<q^*_{\ell m}\Psi_{\ell m}\right> }=\frac{\sum_{\ell=2, m}\oint_{\rm cyc}\dot{q}^*_{\ell m}\Psi_{\ell m} }{\sum_{{\rm osc},\ell =2,m} 2\pi\,q^*_{\ell m}\Psi_{\ell m}}.
\end{eqnarray}
As in the case of a set of decoupled forced non-resonant weakly damped harmonic 
oscillators, a single $Q$ for the combined system only makes sense if
there is a single tidal forcing frequency.  The time derivative $\dot{q}^*_{\ell m}$
enforces that the $\sum_{\ell m}$ in the numerator runs over only the oscillating 
components of the tidal potential and the corresponding tidal response.
Expand into Fourier harmonics and write 

\begin{eqnarray}
\frac{1}{Q}&=&\frac{\sum_{\ell=2, m k}\delta_{\ell m k} \abs{\Psi_{\ell m k}}^2/\varpi^2_{\ell}}{\sum_{{\rm osc},\ell=2, m k} \abs{\Psi_{\ell m k}^2}/
\varpi^2_{\ell}}.
\label{e: tidal_Q}
\end{eqnarray}
Each individual phase lag $\delta_{\ell m k}=\sigma_{mk}\tau_{\ell}$ are equal 
to one another in the event that their respective lag times as equal to one another
as well.  If this is the case, we may then write
\be
\frac{1}{Q}=\delta_{\ell mk}=\delta.
\ee

Comparison of eq. \ref{e: tidal_Q} with eq. \ref{e: Q_phase} shows a
correspondence between
a set of non-resonantly forced, weakly damped harmonic oscillators $(X_n,F_n)$ with
the tidal problem $(q_{\ell m},\Psi_{\ell m})$ in the equilibrium weak-friction limit.  Though the general tidal 
problem is not equivalent to a set of decoupled harmonic oscillators with constant
time lags, their respective expressions for $Q$ are equivalent under
these limited conditions.  
That is, a single tidal $Q$ that represents a single phase lag only holds when $e\ll1$ and $\Omega=n$ for circularization 
or $e=0$ during synchronization.  In addition, the time lag for each of the 
$m$ components of the tidal force of degree $\ell=2$ must be equal to one another.  From this, we arrive at the following conclusion: {\it a single $Q$ can serve to parameterize the level of tidal dissipation 
only if there is a single forcing frequency $\sigma$ and a single constant lag time
$\tau$.}  That is, parameterization of tidal dissipation with a constant $\tau$ is applicable in a more general class of orbital conditions in comparison to 
parameterization with a tidal $Q$.

\end{document}